\documentstyle[preprint,aps,prb,psfig]{revtex}
\tightenlines

\begin{document}

\ifpreprintsty\else
\twocolumn[\hsize\textwidth
\columnwidth\hsize\csname@twocolumnfalse\endcsname       \fi    

%\preprint{  }  \title  {How to tailor many-body interactions in quantum dots}
\title{
Field-induced Coulomb coupling in semiconductor macroatoms: \\
Application to ``single-electron'' quantum devices
}
\author{
Irene D'Amico$^{1,2}$ and Fausto Rossi$^{1,2,3}$
}
\address{$^1$ Istituto Nazionale per la Fisica della Materia (INFM)}
\address{$^2$ Institute for Scientific Interchange (ISI), 
Villa Gualino, Viale Settimio Severo 65, I-10133 Torino, Italy}
\address{$^3$ Dipartimento di Fisica, Politecnico di Torino,
Corso Duca degli Abruzzi 24 \\
I-10129 Torino, Italy}

\date{\today} 
\maketitle

\begin{abstract}

A novel approach for the control of exciton-exciton Coulomb coupling in 
semiconductor macroatoms/molecules is proposed. We show that by applying  
properly tailored external fields, we can induce ---or significantly 
reinforce--- excitonic dipoles, which in turn allows to control and magnify 
intra- as well as inter-dot few-exciton effects. 
Such dipole-dipole interaction mechanism will be accounted for within a 
simple analytical model, which is found to be in good agreement with 
fully three-dimensional calculations.
The proposed approach may play an important role for the design and 
realization of fully-optical quantum gates as well
as ultrafast optical switches.

  \end{abstract}

 \ifpreprintsty\else\vskip1pc]\fi

\narrowtext

In recent years semiconductor nanostructures have gathered enormous 
attention.\cite{Shah}
In particular, the ability of building {\it zero-dimensional} systems, 
such as semiconductor quantum dots (QD's),\cite{gen} led to
 a technological revolution: QD's
applications in fact range
from laser emitters\cite{laser}
to charge-storage devices,\cite{stor} from fluorescent biological 
markers\cite{bio} to quantum information processing devices.\cite{QIP,PRL}
With QD's, the flexibility  in controlling 
carrier densities reaches its extreme:
  it is possible to inject in a QD even just a single electron\cite{Coulblock} 
or  a single 
exciton.\cite{excit,Hawry} 
Their discrete  energy spectrum, on the other side, results 
in a rich optical response and in a weak interaction of the quantized 
carrier system with environmental degrees of freedom (phonons, plasmons, etc).
At the same time, their reduced spatial extension ---up to few 
nanometers--- leads to a significant increase of 
Coulomb interactions among carriers. This, together with the granular 
nature of charge injection/photogeneration previously mentioned, leads to 
pronounced few-particle effects.

In this Letter we shall show how an external electric field can be used
as a simple, effective way to tailor few-exciton interactions 
in single as well as coupled QD structures.
This fine-tuning possibility may
allow for important technological applications, like {\it all-optical} 
quantum gates\cite{PRL} and ultrafast optical switches.\cite{UOS} 
As we shall see, the proposed field-induced effects may be easily understood 
and quantitatively estimated using the simple analytical model described 
below.\cite{model}

Let's consider first a single QD structure.
In the usual situation, the electron and hole charge distributions 
corresponding to the same excitonic state are spatially superimposed. 
Assuming that the typical lengths associated to the confining potential for 
electron and hole are the same, it is possible to 
show that there is no  net Coulomb interaction  among excitons 
belonging to the same ``shell".\cite{Hawry}
In particular, then, if we consider two excitons
with opposite spins in the lower energy state, the associated 
biexcitonic shift\cite{BS} is equal to zero.
Let's apply now a constant electric field. This
 will pull apart charges of  opposite sign, creating  an
electrical dipole for  each exciton. It follows that the net Coulomb 
interaction among excitons is now different from zero even in the same shell. 
In this way  an external electric field, removing part of the system 
symmetries, allows us to {\it turn on} at will and to {\it tune} 
exciton-exciton interactions. 

A similar argument can be applied to the interaction between excitons 
in a coupled QD structure, i.e., to a semiconductor macromolecule.
Even if charge distributions for electrons and holes in the same shell are
somewhat different, the interaction between excitons sitting in different 
and far enough QD's will be negligible. 
If we now apply a constant in-plane electric field (see below),
we create again electron-hole dipoles inside each
QD. Therefore, the  polarized excitons will now interact with a strength 
that is roughly proportional to the square of
the field-induced excitonic dipole.
Once more, the presence of the field is found to turn on exciton-exciton
interactions, thus allowing for the formation of {\it tunable bonds} 
between QD's, i.e., {\it artificial macromolecules}.

The external field can be even used  to {\it turn off} Coulomb
interactions.
Let's consider for simplicity a single exciton in a QD: what happens
if we keep increasing the external electric field?
After the regime of strong electron-hole Coulomb interaction just described,
the electric  field will start to dominate, and the Coulomb correlation 
becomes a mere perturbation, up to  the point in which the electron-hole charge 
separation is so large that
the two particles can be 
described as non-interacting.

In order to test the viability of the proposed scheme, we have performed a 
realistic  calculation\cite{EDA} of 
field-induced exciton-exciton interaction in a GaAs-based coupled QD 
structure.
The upper inset of Fig.~\ref{fig1} shows the electron and hole particle 
distributions 
corresponding to the excitonic ground state along the field direction: 
we can clearly recognize the field-induced electron-hole charge 
separation. Figure 
\ref{fig1} reports the dipole-dipole coupling energy, i.e., the 
biexcitonic shift,\cite{BS} as a function of the in-plane field $E$. 
Here, our exact calculation (squares) is compared to the result of the
model (solid curve) described below. 
As we can see, we obtain energy shifts of the order of a few meV, 
fully compatible with the typical resolution of current ultrafast 
spectroscopy.\cite{Shah} 
We underline that, in order to minimize the loss
 of oscillator strength due to the field-induced charge separation,
a careful optimization 
of the system parameters is needed.

We shall now show that the field-induced effects previously introduced 
can be described in terms of a simple analytical model.
To this purpose, the QD carrier confinement  
along the growth ($z$) direction can be modeled 
as a  {\it narrow} harmonic potential (or as a 
square box)  $V(z)$; the confinement
in the 
QD ($\vec{r}$) plane is described as a two-dimensional (2D) parabolic potential.
By denoting with $\vec{E}$ the in-plane electric field, 
the single-exciton Hamiltonian will then be
\begin{eqnarray} 
{\bf H} &=& \sum_{i=e,h} \left[{{\bf p}_i^2\over 2m_i}  +
{1\over 2} m_i\omega_i^2 \left|\vec{r_i} \pm \vec{d}_i\right|^2 +V_i(z_i)
\right] \nonumber \\
& & -{e^2\over\epsilon 
 \sqrt{|\vec{r_e}-\vec{r_h} |^2 + |z_e-z_h|^2}
}
\ ,
\label{genH}
\end{eqnarray} 
where the $\pm$ sign and the subscripts $e$ and $h$
 refer, respectively, to electron and hole.
Here, 
$\vec{d}_i = {e\vec{E} \over m_i\omega_i^2}$ is the single-particle charge 
displacement induced by the field.
We want to show that, under suitable  conditions,
Eq.~(\ref{genH}) can be analytically solved and all the important 
quantities can be 
easily estimated  with a good degree of accuracy.

Due to the strong single-particle confinement along the $z$ direction, 
we will approximate $|z_e-z_h|^2$ with its average value $\lambda_z^2$.
It is then possible to separate the Hamiltonian (\ref{genH}) as
${\bf H} = {\bf H}_\parallel(\{\vec{r'_i}\}) + {\bf H}_\perp(z_e) + 
{\bf H}_\perp(z_h)$
where
${\bf H}_\perp(z_i) = {p_{z_i}^2\over 2m_i} +V_i(z_i)$
is the single-particle Hamiltonian along the growth direction 
---exactly solvable for the case of a parabolic potential as well as of 
an infinite-height  square well.
By further defining the center of mass (CM) and  relative coordinates 
$\vec{R} = {m_e (\vec{r_e}+\vec{d}_e) + m_h (\vec{r_h}-\vec{d}_h)
 \over M}$ ($M = m_e + m_h$)
and
$\vec{r} = \vec{r_h} - \vec{r_e}$, 
the in-plane Hamiltonian ${\bf H}_\parallel(\{\vec{r'_i}\})$ becomes
\begin{eqnarray}
{\bf H}_\parallel(\vec{R},\vec{r}) 
&=& 
{P^2\over 2M}  + {1\over 2} M\omega_R^2 R^2 +
 {p\over 2\mu } + {1\over 2} \mu\omega_r^2 |\vec{d}-\vec{r} |^2 
 \nonumber \\ 
& & + \mu (\omega_e^2 - \omega_h^2) 
\vec{R}\cdot (\vec{d}-\vec{r})
- {e^2\over\epsilon \sqrt{r^2+\lambda_z^2}} \ ,
\label{parH2}
\end{eqnarray}
where $\mu = {m_em_h \over M}$ is the reduced mass,
$\omega_R^2 = {\omega_e^2 + \omega_h^2 \over 2} (1 + \Delta)$,
$\omega_r^2 = {\omega_e^2 + \omega_h^2 \over 2} (1 -\Delta)$,  
$\Delta = {m_e-m_h \over M} 
{\omega_e^2-\omega_h^2 \over \omega_e^2+\omega_h^2}$ and $\vec{d} = \vec{d}_e+\vec{d}_h$ denotes the total electron-hole 
in-plane displacement.
 
In the limit 
${\omega_e^2-\omega_h^2 \over \omega_e^2+\omega_h^2} \ll 1$, 
the two coordinates are only weakly coupled, and 
the Schr\"odinger equation associated to the CM coordinate $\vec{R}$ is 
exactly solvable.
In the general case we will concentrate on the ground state, though 
the generalization to higher states is straightforward.
We  can approximate the ground state of 
${\bf H}_\parallel$ as
$
\Psi(\vec{r},\vec{R}) =
\Psi_x(x) {1\over(\lambda_r^2\pi)^{1/4}}
e^{-{y^2 \over 2 \lambda_r^2}} 
{1\over(\lambda_R^2\pi)^{1/2}}e^{-{R^2 \over 2 \lambda_R^2}}$,
where $x$ and $y$ denote, respectively, the components of $\vec{r}$ 
parallel and perpendicular to the field $\vec{E}$,
$\lambda_r = \sqrt{{\hbar \over \mu\omega_r}}$ 
and $\lambda_R = \sqrt{{\hbar \over M\omega_R}}$.
By averaging ${\bf H}_\parallel$ over $\Psi(\vec{r},\vec{R})$, 
we obtain the effective Hamiltonian 
${\bf H}_{eff} = {1 \over 2}\hbar\omega_r +\hbar\omega_R
+ {p_x^2 \over 2\mu} + V_{eff}(x)$,
characterized by the effective potential 
\begin{equation}
V_{eff}(x) = {1\over 2} \mu  \omega_r^2
(x-d)^2 + V_C\left({x^2+\lambda_z^2 \over 2\lambda_r^2}\right)
\label{effV}
\end{equation}
with
$V_C(u) = - {e^2 \over\epsilon \sqrt{\pi} \lambda_r} 
e^{u} K_0(u)$,
$K_0$ being the zero-order Bessel function.

If we are interested in the low-energy states, 
we can approximate $V_{eff}$ around its minimum $V_0$ with a 
parabolic potential, i.e.,
$V_{eff}(x) \approx V_0+{1\over 2} \mu\tilde{\omega}^2(x-x_0)^2$.
Within such approximation scheme, the eigenvalues and eigenfunctions of 
${\bf H}_{eff}$ can be evaluated analytically.
As already pointed out, if the external field is  strong enough (and we will
quantitatively  define ``enough'' later),  the Coulomb 
attraction between electron and hole 
can be considered as a perturbation. In this regime its main effect
is to reduce the electron-hole displacement from $d$ to $x_0$.
For intermediate and strong fields, we can then write the effective 
displacement as
$x_0= d-\Delta x$, with $\Delta x\ll d$.
In this regime the following analytical expression for $\Delta x$ is 
obtained:
\begin{equation}\label{delx}
{\Delta x\over d} = -{\lambda_r\over a^*}{\exp(\xi)\over\sqrt{\pi}}
{\Delta K \over 1-{\lambda_r\over a^*}{\exp(\xi)\over\sqrt{\pi}}
\left[{d^2\over\lambda_r^2}A(\Delta K,K_1)+\Delta K\right]},
\end{equation}
with $\xi = (d^2+\lambda_z^2)/2\lambda_r^2$, $K_1$ the first-order Bessel 
function, $\Delta K=K_0(\xi)-K_1(\xi)$, $A(\Delta K,K_1)=2\Delta K+
{K_1(\xi)\over \xi}$ and $a^*={\hbar^2\epsilon\over\mu e^2}$  the reduced Bohr
radius. Notice that the prefactor ${\lambda_r\over a^*}$ is a measure of the 
system confinement.
In a similar way, setting $\tilde{\omega}=\omega_r+\Delta\omega$ in $\mu\tilde{\omega}=
d^2 V_{eff}/dx^2|_{x_0}$,
 we can calculate  the effect of the Coulomb 
attraction on the potential shape.
%\begin{eqnarray}
%{\Delta\omega\over \omega_r}&=&
%-{\lambda_r\over a^*}{\exp(\xi)\over 2\sqrt{\pi}}
%({d^2\over\lambda_r^2}A(\Delta K,K_1)+\Delta K\nonumber \\
%&-&{\Delta x\over d}
%{d^2\over\lambda_r^2}\left\{{d^2\over\lambda_r^2}\left[2A(\Delta K,K_1)-{1\over\xi}
%(\Delta K+2{K_1(\xi)\over \xi})\right]+3A(\Delta K,K_1)\right\}).
%\end{eqnarray}
The result is cumbersome and not particularly highlighting; 
here we will report only 
the important limit
${\lambda_r^2\over d^2}\ll 1$ (high fields), in which
$ {\Delta\omega\over \omega_r}=-{\Delta x\over d}\propto -{\lambda_r\over a^*}
{\lambda_r^3\over d^3}$.
The condition $\Delta x/d\stackrel{<}{\sim}20\%$ 
quantitatively defines the ``intermediate and 
strong'' electric field regime.
It is easy to show that, in the regime of interest, the correction 
on the wave function due to
 $\Delta \omega/
\omega_r$ is negligible with respect to the correction given by the shift
$\Delta x/d$. 

Based on the analytical model proposed so far, we have investigated the 
biexcitonic shift previously discussed (see Fig.~\ref{fig1}) 
more specifically, we have approximated the biexcitonic ground state 
as the product of two excitonic wavefunctions sitting in 
different dots. The latter, in turn, are taken as products of the in-plane 
wavefunctions  times the ground-state wavefunctions along the
$z$ direction.
 The desired biexcitonic shift $\Delta \epsilon$ is then obtained 
averaging the corresponding two-exciton Hamiltonian over such factorized 
ground state.
Within this approximation, $\Delta \epsilon$ can be reduced to
 an easy-to-calculate sum of, at 
most, two-dimensional integrals.
In the corresponding validity region 
the estimate provided by the model is accurate: 
Figure \ref{fig1} shows the difference between the exact results (squares),
the approximate results (solid curve) and the results obtained neglecting 
Coulomb correlation completely (dotted line). 
The dashed curve shows the approximate results 
obtained setting $\Delta\omega/\omega_r=0$: as anticipated, this correction 
is generally negligible. The lower inset presents the behavior of $\Delta x/d$ and
$\Delta\omega/\omega_r=0$ with respect to the external field 
$\vec{E}$.
We stress that the proposed model allows for
a  quick scan of the whole parameter space, useful especially when
it is complex to determine the correct 
operative region and the
 exact numerical calculation  requires  a long computational 
time.\cite{PRL}

In conclusion, we have discussed how an external electric field can be used
to turn on and off exciton-exciton interactions in a QD system and we have 
provided a simple analytical model to calculate its  main properties. 
The proposed strategy can play an important role in the design and 
optimization of semiconductor-based quantum devices, like ultrafast optical 
switches, single-electron devices, and quantum-information processors. 

\medskip

We are grateful to Eliana Biolatti, Rita Iotti, and Paolo Zanardi for 
stimulating and fruitful discussions.
This work has been supported in part by the European Commission through the
Research Project {\it SQID} within the {\it Future and Emerging 
Technologies (FET)} programme.

\begin{figure}
\psfig{figure=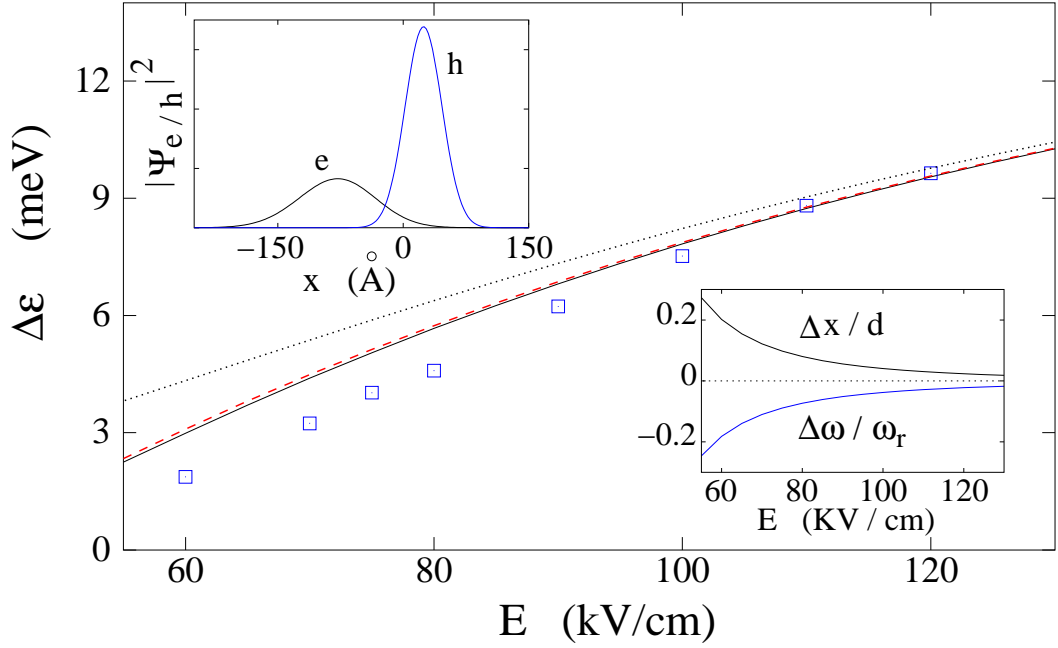,angle=-90,width=1.05\columnwidth}
\caption{Biexcitonic shift $\Delta \epsilon$ vs external field $E$ for a system 
characterized by the parameters $m_e=0.067 m_0$,  $m_h=0.34 m_0$ ($m_0$ the 
metallic
electron mass), $\hbar\omega_e=30~ meV$ and $\hbar\omega_h=24~ meV$.
The squares indicate the results of the fully three dimensional calculation,
the solid line represents the results of the proposed model, the dotted line
is the result obtained when Coulomb correlations are completely neglected and
the dashed line corresponds to setting $\Delta\omega =0$ in the model.
The upper inset shows the electron and hole particle density corresponding to 
the excitonic ground state along the field direction; the lower inset presents 
the behavior of the two key quantities $\Delta\omega/\omega_r$ and $\Delta x/d$
in respect to the external field.       
}
\label{fig1}
\end{figure}

%\begin{figure}
%\psfig{figure=../V_effcomp.ps,width=.95\columnwidth}
%\caption{
%}
%\label{fig2}
%\end{figure}


\begin{references}
\bibitem{model}Similar models have been proposed for different situations
in Ref. \protect\onlinecite{gen}\protect
\bibitem{BS}
The biexcitonic shift is defined as the difference between excitonic 
and biexcitonic optical transition energies.\protect\cite{gen}\protect
\bibitem{EDA}
Our fully three-dimensional calculation of the biexcitonic ground state is 
based on the direct-diagonalization approach proposed in 
Ref.~\protect\onlinecite{PRL}\protect.
\end{references}
\end{document}